\documentclass[aps,prb,reprint,twocolumn,superscriptaddress,showpacs]{revtex4-1}

\usepackage{graphicx}
\usepackage[pdftex,colorlinks=true,linkcolor=blue,citecolor=blue,urlcolor=blue,filecolor=blue]{hyperref}

\begin{document}

\title{Direct correlation of crystal structure and optical properties
in wurtzite/zinc-blende GaAs nanowire heterostructures}

\author{Martin Hei\ss} \affiliation{Walter Schottky Institut and
Physik Department, Technische Universit\"at M\"unchen, Am
Coulombwall~3, D-85748 Garching, Germany} \affiliation{Laboratoire des
Mat\'{e}riaux Semiconducteurs, Institut des Mat\'{e}riaux, Ecole
Polytechnique F\'{e}d\'{e}rale de Lausanne, Switzerland}

\author{Sonia Conesa-Boj} \affiliation{Laboratoire des Mat\'{e}riaux
Semiconducteurs, Institut des Mat\'{e}riaux, Ecole Polytechnique
F\'{e}d\'{e}rale de Lausanne, Switzerland} \affiliation{Departament
d'Electr\`{o}nica,Universitat de Barcelona, E-08028 Barcelona, CAT,
Spain}

\author{Jun Ren} \affiliation{Laboratory for Multiscale Modeling of
Materials, Institut des Mat\'{e}riaux, Ecole Polytechnique
F\'{e}d\'{e}rale de Lausanne, Switzerland}

\author{Hsiang-Han Tseng} \affiliation{Laboratory for Multiscale
Modeling of Materials, Institut des Mat\'{e}riaux, Ecole Polytechnique
F\'{e}d\'{e}rale de Lausanne, Switzerland}

\author{Adam Gali} \affiliation{Department of Atomic Physics, Budapest
University of Technology and Economics, Budafoki \'ut 8., H-1111,
Budapest, Hungary}

\author{Andreas Rudolph} \affiliation{Institute for Experimental and
Applied Physics, University of Regensburg, Universit\"atsstrasse 31,
D-93053 Regensburg, Germany}

\author{Emanuele Uccelli} \affiliation{Walter Schottky Institut and
Physik Department, Technische Universit\"at M\"unchen, Am
Coulombwall~3, D-85748 Garching, Germany} \affiliation{Laboratoire des
Mat\'{e}riaux Semiconducteurs, Institut des Mat\'{e}riaux, Ecole
Polytechnique F\'{e}d\'{e}rale de Lausanne, Switzerland}

\author{Francesca Peir\'{o}} \affiliation{Departament
d'Electr\`{o}nica,Universitat de Barcelona, E-08028 Barcelona, CAT,
Spain}

\author{Joan Ramon Morante} \affiliation{Catalonia Institute for
Energy Research (IREC), E-08019 Barcelona, CAT, Spain}
\affiliation{Departament d'Electr\`{o}nica,Universitat de Barcelona,
E-08028 Barcelona, CAT, Spain}

\author{Dieter Schuh} \affiliation{Institute for Experimental and
Applied Physics, University of Regensburg, Universit\"atsstrasse 31,
D-93053 Regensburg, Germany}

\author{Elisabeth Reiger} \affiliation{Institute for Experimental and
Applied Physics, University of Regensburg, Universit\"atsstrasse 31,
D-93053 Regensburg, Germany}

\author{Efthimios Kaxiras} \affiliation{Laboratory for Multiscale
Modeling of Materials, Institut des Mat\'{e}riaux, Ecole Polytechnique
F\'{e}d\'{e}rale de Lausanne, Switzerland}

\author{Jordi Arbiol} \affiliation{Instituci\'{o} Catalana de Recerca
i Estudis Avan\c{c}ats (ICREA) and Institut de Ci\`{e}ncia de
Materials de Barcelona, CSIC, 08193 Bellaterra, CAT, Spain}

\author{Anna Fontcuberta i Morral}
\email{anna.fontcuberta-morral@epfl.ch}

\affiliation{Walter Schottky Institut and Physik Department,
Technische Universit\"at M\"unchen, Am Coulombwall~3, D-85748
Garching, Germany} \affiliation{Laboratoire des Mat\'{e}riaux
Semiconducteurs, Institut des Mat\'{e}riaux, Ecole Polytechnique
F\'{e}d\'{e}rale de Lausanne, Switzerland}

\date{\today}

\begin{abstract} A novel method for the direct correlation at the
nanoscale of structural and optical properties of single GaAs
nanowires is reported. Nanowires consisting of 100\% wurtzite and
nanowires presenting zinc-blende/wurtzite polytypism are investigated
by photoluminescence spectroscopy and transmission electron
microscopy. The photoluminescence of wurtzite GaAs is consistent with
a band gap of 1.5\,eV. In the polytypic nanowires, it is shown that
the regions that are predominantly composed of either zinc-blende or
wurtzite phase show photoluminescence emission close to the bulk GaAs
band gap, while regions composed of a nonperiodic superlattice of
wurtzite and zinc-blende phases exhibit a redshift of the
photoluminescence spectra as low as 1.455\,eV. The dimensions of the
quantum heterostructures are correlated with the light emission,
allowing us to determine the band alignment between these two
crystalline phases. Our first-principles electronic structure
calculations within density functional theory, employing a
hybrid-exchange functional, predict band offsets and effective masses
in good agreement with experimental results.

\end{abstract}


\maketitle

\section{Introduction}

Semiconductor nanowires are attracting increasing interest because of
their exciting optical and electronic properties and the possibility
of synthesizing them in a controllable
fashion~\cite{Cui2001,ref2,Cui2003,Bjork2002}. They are considered
promising building blocks for the next generation of optical, sensing,
electronic and energy harvesting
devices~\cite{Wang2005,Duan2001,Chan2008,Tian2007}. Their small
diameters, often at distances a few times the inter-atomic spacing in
crystals, and their unique geometries lead to physical properties
which differ significantly from the corresponding bulk material.
Furthermore, nanowires hold the promise of integrating
lattice-mismatched crystals on a single device, which opens new design
possibilities and has the potential to significantly decrease
production costs in solar cell applications\cite{Kayes2005}.

The functionality of nanowires is increased, and their properties are
modified, when radial or axial heterostructures are created within a
single wire\cite{Lauhon2002,Gudiksen2002}. Radial heterostructures can
be produced by growing different materials on the side facets of the
nanowires \cite{prismatic1,prismatic2}, while axial heterostructures
correspond generally to a variation in the chemical composition along
the growth axis\cite{Bjork2002,ref2,indium1}. Recently, a second type
of axial heterostructure has attracted attention, where the
crystalline structure varies along the growth axis while the chemical
composition is conserved. Typically, the structure is changed from the
cubic zinc-blende to the hexagonal wurtzite
structure~\cite{wzzb0,wzzb1,wzzb2}. This change of crystalline
structure has been observed in some bulk materials such as SiC and
GaN~\cite{Park1994,Strite1992}. In phosphides and arsenides, it has
been shown that the appearance of wurtzite phase is restricted to the
nanowire form and is never observed in the bulk \cite{ref22}. The
change in crystal structure is accompanied by changes in the
electronic structure; for example, earlier theoretical work predicted
a type II band alignment between the zinc-blende and wurtzite phases
in III-V semiconductors \cite{Murayama1994}.

\begin{figure*} \includegraphics[width=\textwidth]{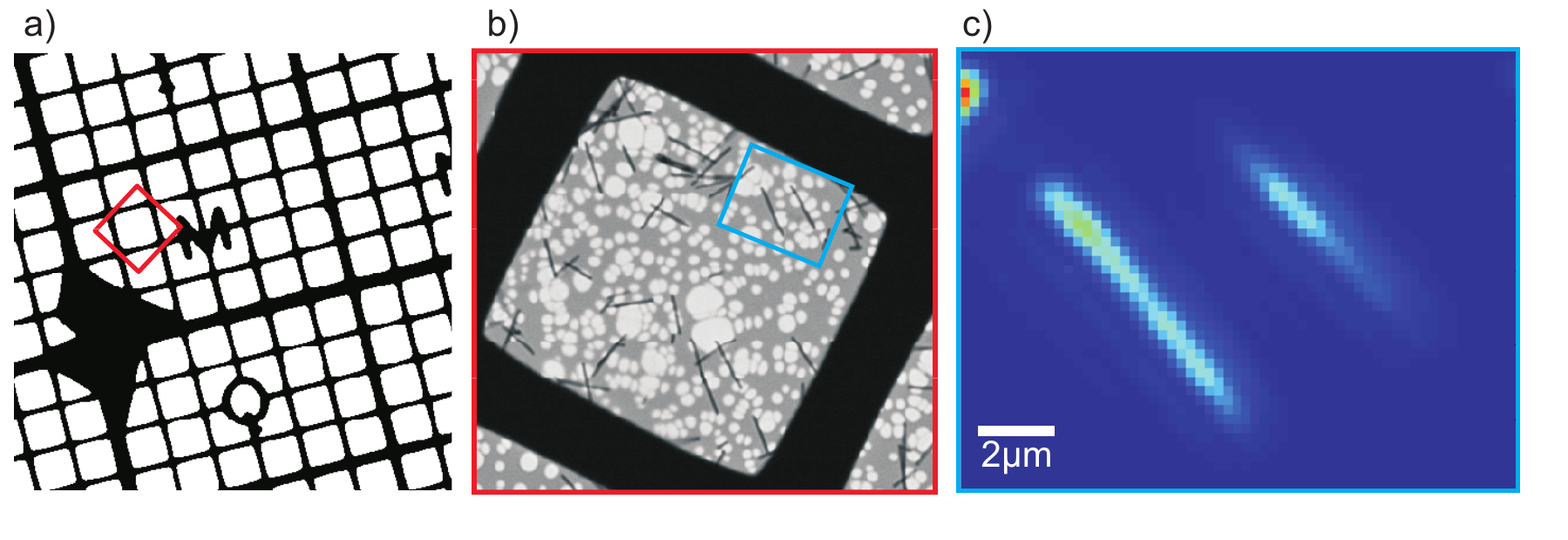}
\caption{\label{figure1} Overview of the measurement process: a)
Optical microscopy image of a TEM copper grid with marker symbols used
for the studies. b) Detail of TEM grid showing the nanowires
transferred to the holey carbon film. c) Color plot of integrated
photoluminescence signal at 4.2 K from the area of the TEM grid marked
by the blue box in b)} \end{figure*}

Recently, type II band alignment in InP nanowires~\cite{nlbao,pemasiri2009} has been
demonstrated allowing the formation of crystal phase qantum dots~\cite{Akopian2010}.
Meanwhile the band structure of wurtzite GaAs has been a subject of controversy in the last few
years. Recent works on the luminescence of GaAs nanowires with a high
percentage of wurtzite phase show a large variety of results. Recently
luminescence of wurtzite GaAs in the range of 1.53-1.54\,eV has been
reported \cite{hoang:133105}. Previously, Martelli \textit{et al.} had
reported an emission at 1.522\,eV in wurtzite GaAs nanostructures
\cite{Martelli2007}. Meanwhile Moewe \textit{et al.} reported an
emission at 10\,meV below the zinc-blende GaAs band
gap\cite{Moewe2008}. Additionally, our own results presenting a
mixture of wurtzite and zinc-blende phases in nanowires are consistent
with a type II band alignment and smaller band gap~\cite{spirkoska-2009}. Indeed in nanowires presenting 30\% and 70\% wurtzite phase we have observed extremely sharp peaks 1.43 and 1.515\,eV~\cite{spirkoska-2009}. The spectra could be explained by the existence of quantum heterostructures forming a type II band alignment~\cite{spirkoska-2009}. Additionally similar results have been obtained by other groups~\cite{Akopian2010}. We believe that the apparent contradictions of
some of these literature results can be related to the fact that the
optical and structural characterizations were not performed on the
same nanowire. This limits the possibilities to account for
variability in the structure of different nanowires from one growth.
In this work we will perform a direct correlation between the
structural and optical properties in order to overcome this
limitation.

To the best of our knowledge, a direct correlation between the optical
properties and the structural sequence in the polytypism, which
requires both functional and structural characterization on the the
same nanowire, has not yet been demonstrated~\cite{Ropers2007}. An initial effort in
this direction was carried out by Bao \textit{et al.}~\cite{nlbao} who
looked at the shift of the luminescence band as a function of the
excitation power for highly twinned and twin-free zinc-blende
structures in InP nanowires. Arbiol \textit{et al.}~\cite{arbiol2009}
demonstrated by in-situ local electron energy loss spectroscopy (EELS)
measurements on [111] zinc-blende inclusions in [0001] wurtzite
nanowires in GaN the variation in the local density of states close to
the band gap. Finally in different type of heterostructures,
transmission electron microscopy (TEM) based cathodoluminescence was
used to characterize GaN/InGaN/AlGaN nanowire
heterostructures~\cite{lim-2009}. This technique enabled the direct
correlation between the structural quality and and the carrier
recombination characteristics in InGaN quantum wells~\cite{lim-2009}.

In this work, we provide a study that demonstrates direct correlation
between the optical properties and structure at the atomic scale of
GaAs nanowires formed by either pure wurtzite-GaAs or a combination of
various thicknesses of the zinc-blende and wurtzite phases. This
enables us to elucidate the band alignment between the two crystalline
phases and represents an important step towards structural band gap
engineering.

\section{Experimental Details}

The nanowires were grown in a Gen II Molecular Beam Epitaxy (MBE)
system. Wurtzite GaAs nanowires were grown by using gold as a
catalyst, while the polytypic GaAs nanowires were obtained by the
gold-free gallium-assisted growth method. In order to avoid
cross-contamination because of the use of gold, the two different
samples were fabricated in separate MBE systems. The nanowires with
$\sim$98-100\% wurtzite GaAs were grown on GaAs
(1$\overline{\text{1}}$1)B substrates at a growth temperature of
540\,$^\circ C$ under a As$_4$ Beam flux of $1.2\cdot10^{-6}$\,mbar at
a Ga rate equivalent to a planar growth of 0.8\,\AA/s. The growth time
was 7500 s. The nucleation and growth followed the Vapor-Liquid-Solid
mechanism, with Au as catalyst\cite{Wagner1964}. Details on the growth
procedure are described in \cite{Rudolph2009}. After the axial growth
of the nanowires the growth parameters were changed to conditions
suitable for planar growth and the nanowires were passivated by an
epitaxial prismatic shell of AlGaAs/GaAs material \cite{prismatic2,
prismatic1}. The 2D equivalent amount grown during capping was 60\,nm
AlGaAs followed by 30\,nm GaAs.

Wurtzite/zinc-blende polytypic nanowires were obtained on
(1$\overline{\text{1}}$1)B GaAs substrates previously coated with
20\,nm SiO$_2$. The growth temperature was $630\,^\circ$C and the As
beam flux at the beginning of the growth process $8.8\times
10^{-7}$\,mbar at a Ga rate equivalent to a planar growth of
0.22\,\AA/s. For this sample, the As beam flux increased slightly up
to approx. $1\times 10^{-6}$\,mbar during the growth. The nanowires
were capped by an epitaxial shell of
AlGaAs/GaAs~\cite{prismatic2,prismatic1}.

\begin{figure} \includegraphics[width=\columnwidth]{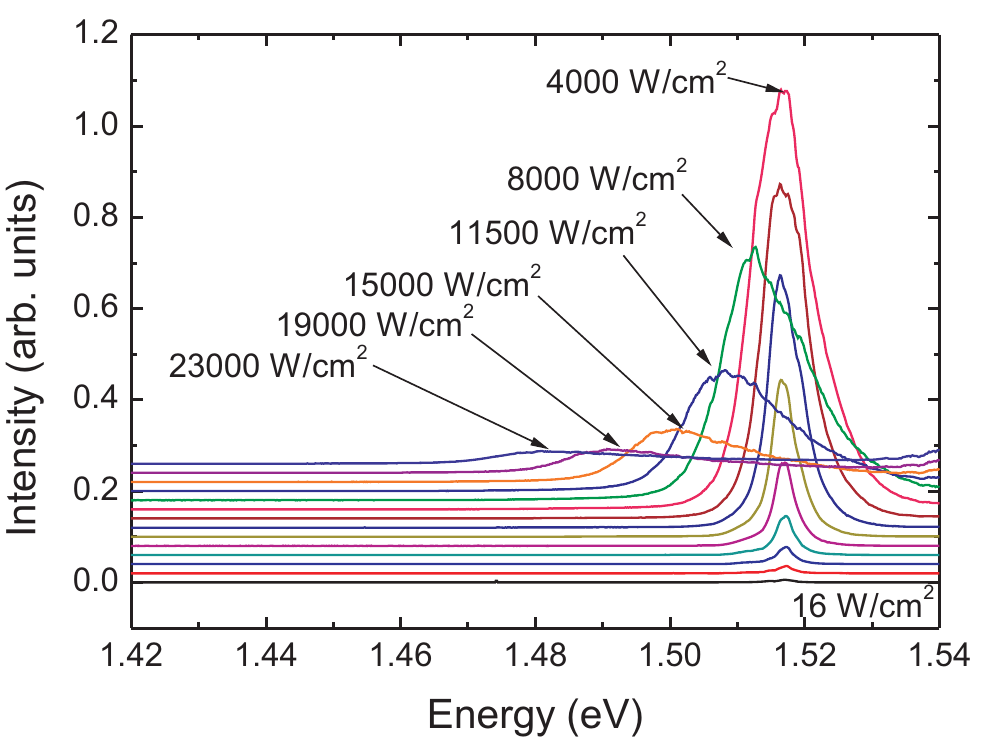}
\caption{\label{pltemheating} Excitation power dependence of a typical
capped zinc-blende GaAs nanowire, dispersed on a holey carbon film of
a copper TEM grid. No pronounced heating effects are observed for
excitation power densities up to 4000\,W/cm$^2$. For higher excitation
powers the heating of the sample causes a change in band gap resulting
in a redshift of the photoluminescence transition. For the highest
excitation of 23000\,W/cm$^2$ the redshift corresponds to a
temperature increase from 4.2\,K to 140\,K. This clearly demonstrates
that heating can be neglected for measurements performed at excitation
powers in the order of 10-50\,W/cm$^2$.} \end{figure}

\begin{figure} \includegraphics[width=\columnwidth]{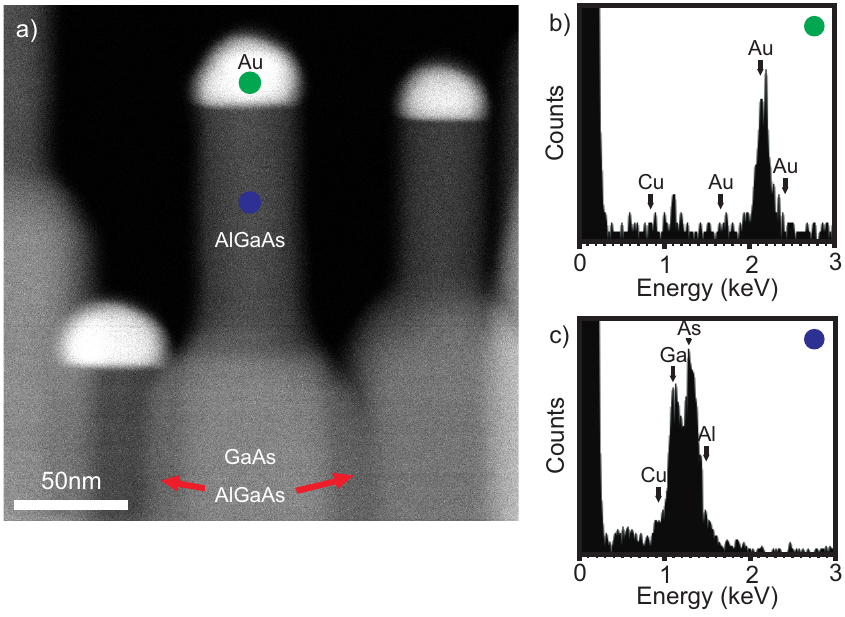}
\caption{\label{wz-neck-edx} a) Typical HAADF STEM micrograph obtained
on a Au-seeded AlGaAs/GaAs NW. b) EDX spectrum taken at the tip of the
nanowire confirming the presence of Au (Green). c) EDX spectrum at the
neck of the nanowire confirming the presence of and AlGaAs on the neck
(percentage of Al is low).} \end{figure}

For the direct correlation experiments, the nanowires were first
transferred onto TEM holey carbon Cu grids. The grids presented
markers in order to facilitate the localization of the nanowires in
the multiple measurements, see Fig.\,\ref{figure1}a-b. For the optical
characterization, the grids were fixed freely suspended to a sample
holder mounted in a confocal micro-photoluminescence ($\mu$PL) setup.
The sample stick was then immersed in a liquid helium bath cryostat.
In order to ensure a good thermal dissipation during the optical
measurements, the sample space was filled with a helium exchange gas
at a pressure of $\sim$\,5 mbar. The $\mathrm{\mu PL}$ measurements
were carried out at a temperature of 4.2\,K. The excitation sources
were a semiconductor laser diode emitting at $780\pm10$\,nm or a
helium neon laser emitting at 632.8\,nm. The laser was focused to a
diffraction limited spot of about 0.8\,$\mu$m in diameter. This
diffraction limit of light of the confocal microscopy is also the
limiting factor for the overall resolution of the correlation
technique. Typical excitation power densities were in the order of
10-50\,W/cm$^2$, orders of magnitude below the level where significant
heating of the sample could be observed (Fig.\,\ref{pltemheating}).
Indeed, in contrast to similar experiments performed in a helium flow
cryostat~\cite{nlbao}, we did not observe any severe heating effects
resulting from the low heat conductivity of the thin carbon membrane.
In our case, we are able to realize $\mathrm{\mu PL}$ measurements on
nanowires dispersed to a free-standing carbon membrane without the
need of adding supplementary layers on top to allow for thermal
dissipation as it was realized in previous studies \cite{nlbao}. Only
this configuration enables simultaneous high resolution TEM and
$\mathrm{\mu PL}$ measurements on the same single nanowire. The
position of the sample in the in the focal plane of the objective was
verified by acquiring a confocal reflectivity image from the sample
area. The $\mathrm{\mu PL}$ measurements were then acquired by
scanning an area of 20\,$\mu$m/side with a step of 250\,nm and
aquiring a spectrum at each position, see Fig.\,\ref{figure1}c. The
spatial dependence of PL spectra along the nanowire axis was obtained
by plotting the suitable cross-section of this dataset.

Right after $\mathrm{\mu PL}$ measurements the nanowires were studied
by TEM. A series of connected bright field HRTEM micrographs was
obtained from the bottom to the top of the nanowire, with the purpose
of obtaining detailed information on the structure. Both high
resolution and low resolution measurements were taken. The micrographs
were taken along the nanowire, ensuring that the ends were overlapping
with the previous/following micrographs. Multiple subsequent
micrographs was necessary to map the whole length. The electron
microscopy (TEM and HRTEM) measurements were carried out in a CM300
LaB6 microscope with a point to point resolution of 0.17\,nm.

\section{Results}

\begin{figure} \includegraphics[width=\columnwidth]{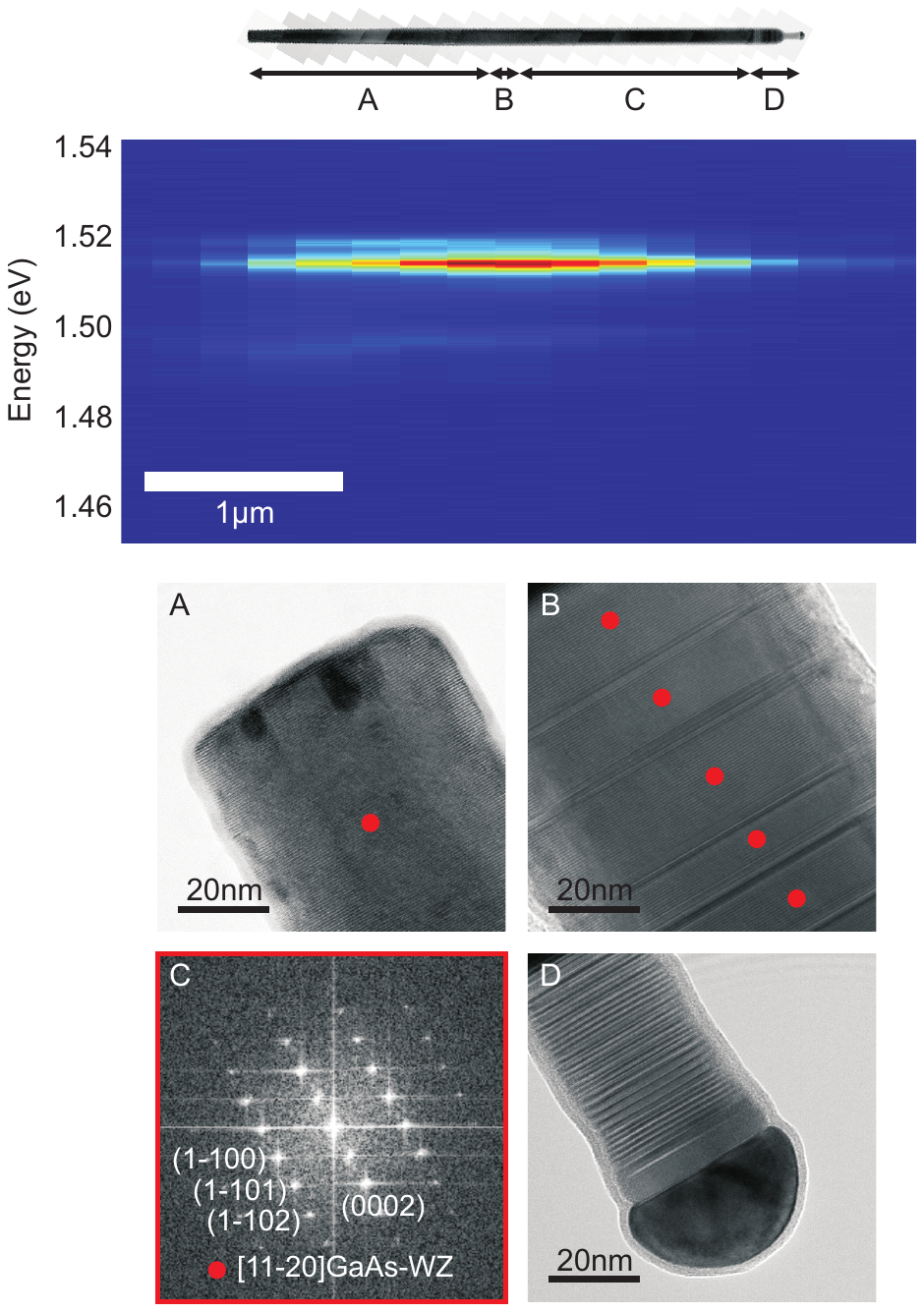}
\caption{\label{wzcorrelation} Mapping of photoluminescence spectra
along the length of the nanowire shown above. The scale bar applies to
both the photoluminescence mapping and the TEM. A,B,D) HRTEM
micrographs showing the crystalline phases at the regions of the
nanowire indicated by the arrows above. Red dots mark wurtzite GaAs
phase. C) Power spectrum of HRTEM micrograph in region C. Note that
micrograph in region D corresponds to the axially grown AlGaAs/GaAs
material during the radial capping forming a thin extension.}
\end{figure}

\begin{figure} \includegraphics[width=\columnwidth]{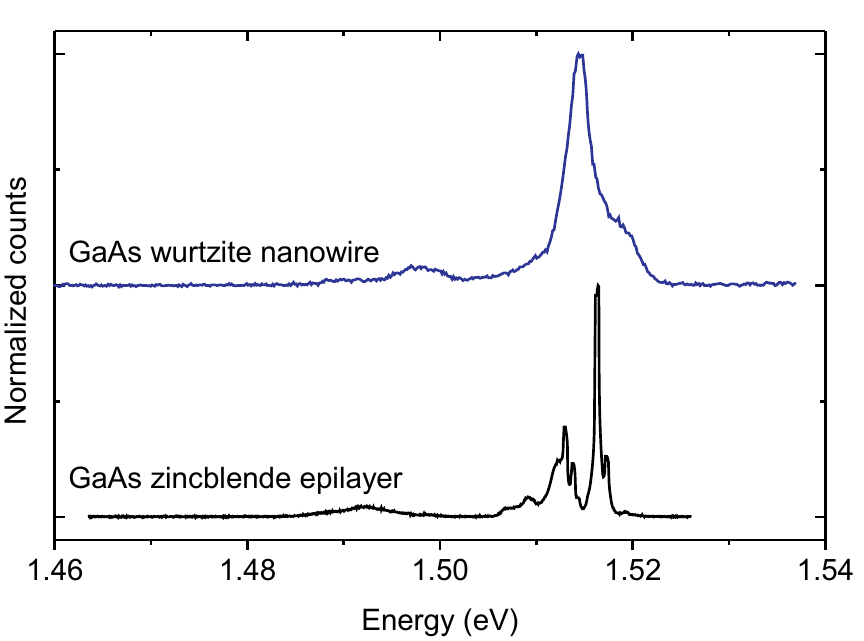}
\caption{\label{wzspectra} Photoluminescence spectrum from nanowire
consisting of pure wurtzite GaAs phase (top); Photoluminescence of a
planar zinc-blende GaAs reference sample (bottom).} \end{figure}

\subsection{Wurtzite GaAs band gap}

We start by presenting results in nanowires consisting of nearly 100\%
wurtzite. In this case, the direct correlation method is less
important, as there is no significant variation on the structure along
the nanowire. The correlation allows us to verify that the
photoluminescence characteristics are correctly attributed to
nanowires composed from wurtzite GaAs.

Fig.\,\ref{wz-neck-edx}a shows a high angle annular dark field (HAADF)
STEM micrograph of several nanowires from the sample. In principle the
HAADF detector directly gives a contrast proportional to the atomic
charge Z of the nuclei in the sample. The contrast also depends on the
sample thickness that is not uniform in the axial extension compared
to the rest of the wire. The image contrast in
Fig.\,\ref{wz-neck-edx}a therefore cannot directly indicate the
chemical composition. For this reason energy dispersive X-ray
spectroscopy (EDX) was performed at the thin extension and the
catalyst droplet to clearly determine the chemical composition.
Fig.\,\ref{wz-neck-edx}b shows the EDX spectrum taken with the
electron beam focused at the tip of the nanowire. The spectral
features clearly demonstrate the presence of the Au catalyst.
Meanwhile Fig.\,\ref{wz-neck-edx}c shows that the thin extension of
the nanowire towards the tip corresponds to axially grown AlGaAs. This
is in clear agreement with the picture that during the radial capping
with the AlGaAs/GaAs shell the nanowire continues to grow axially from
precipitation below the catalyst droplet. Indeed for the case of Au
catalyzed nanowire the metallic catalyst is not affected by the growth
conditions. We can therefore use the size of the nanowire neck to
accurately estimate the size of the nanowire core.

Now we turn to the correlation between HRTEM and spatial dependent PL.
In Fig.\,\ref{wzcorrelation} we have included TEM micrographs
representative of various locations along the nanowire. As shown at
the left end the nanowire consists of pure wurtzite GaAs free of
stacking faults throughout a $1.2\,\mathrm{\mu m}$ long region A
(Fig.\,\ref{wzcorrelation}). In the 150\,nm long region B we can
observe seven stacking faults in the nanowire that is otherwise still
composed of pure wurtzite phase. The following $1\,\mathrm{\mu m}$ of
the nanowire (C) are stacking-fault free wurtzite GaAs material. The
final 250\,nm (D) of the nanowire exhibits a mixture of heavily
twinned zinc-blende and wurtzite phases. This can be attributed to the
stage in which the conditions (mainly As pressure) were changed from
axial to predominantly radial growth. Region D corresponds to the
remaining axial growth occurring during the capping and is composed of
AlGaAs as discussed before. Fig.\,\ref{wzspectra} shows the
typical PL spectra along the approximately $2.8\,\mathrm{\mu m}$ long
nanowire. An emission peak at approx. 1.515\,eV is observed throughout
the length of the nanowire. By exciting the luminescence with an HeNe
laser with a photon energy of 1.96\,eV, we have verified that there is
no emission in the range of energies between 1.52 and 1.63\,eV. This
is in agreement with our previous studies in which the luminescence in
wurtzite/zinc-blende heterostructured nanowires was always observed
below 1.515\,eV \cite{spirkoska-2009}. Furthermore this value is in
agreement with recent theoretical predictions \cite{De2010}.

We have examined at least 4 nanowires with the same structural
characteristics (nearly 100\% wurtzite) and we did not observe any
luminescence at the energies between 1.52 and 1.63\,eV . Additionally,
in order to rule out possible contamination of the nanowires, we have
measured the PL of a $1\,\mathrm{\mu m}$ thick GaAs epilayer. The
spectrum is shown in Fig.\,\ref{wzspectra} The emission is peaked at
1.516\,eV, as expected for excitonic transitions in high purity GaAs.
The additional sharp peaks in the PL spectrum as well as the broad
emission at 1.49\,eV can be attributed to carbon impurities. As a
consequence we can conclude that our measurements are consistent with
an interband transition in the wurtzite GaAs at 1.515\,eV clearly
originated in the core of the nanowire.

Finally, we want to further discuss the relation between the observed
emission at 1.515\,eV and the band gap of bulk wurtzite GaAs. The
diameter of the catalyst droplet and therefore the nanowire core is
approx. 33\,nm (see Fig.\,\ref{wzcorrelation}D). Due to the small
size, the radial confinement of carriers is causing a slight blue
shift with respect to the actual bulk band gap. We have estimated this
confinement energy by simulating\cite{Birner2007} this structure with
nextnano$^3$. For simplicity the calculations are using the
zinc-blende GaAs band parameters. The simulations show a confinement
energy in the range of 13.5-17.8\,meV for diameters between 35-30\,nm.
By correcting the observed emission energy with the estimated radial
confinement energy we obtain an estimated wurtzite GaAs band gap of
1.50\,eV.

\subsection{Wurtzite/zinc-blende GaAs band offsets} 

\begin{figure}
\includegraphics[width=\columnwidth]{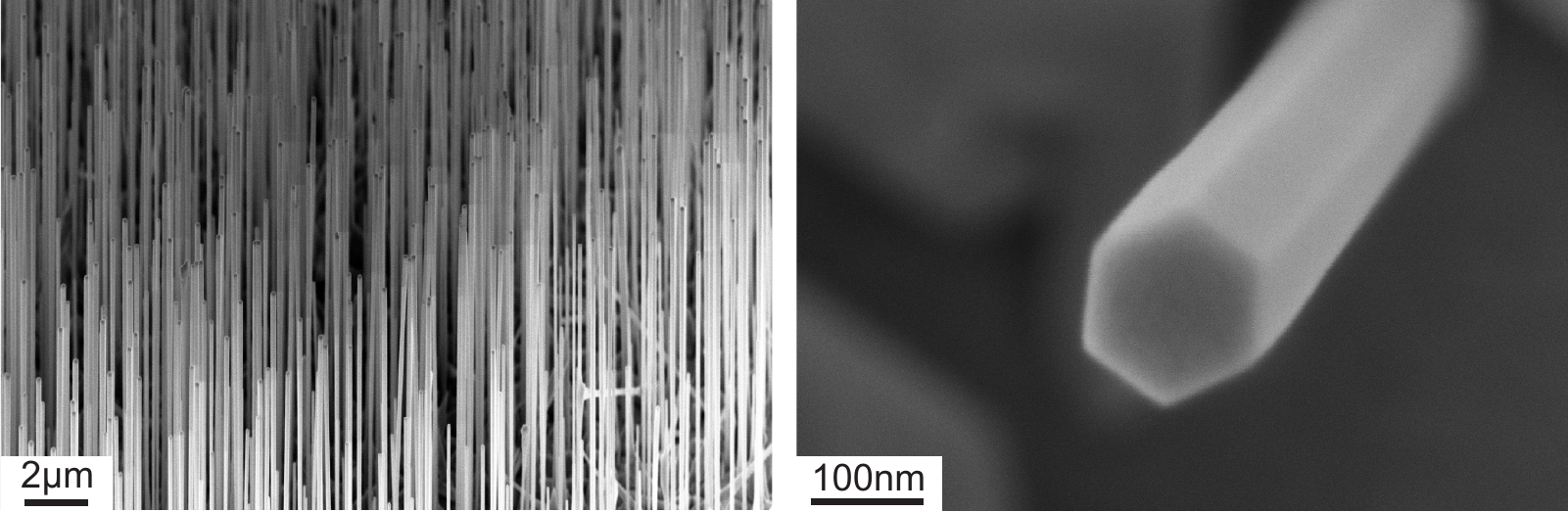}
\caption{\label{semgacatalyzed} Scanning electron micrographs of the
as-grown nanowire sample grown with the Ga assisted growth mechanism.
The nanowires grow perpendicular to the (1$\overline{\text{1}}$1)B
substrate surface.} \end{figure}

\begin{figure*} \includegraphics[width=\textwidth]{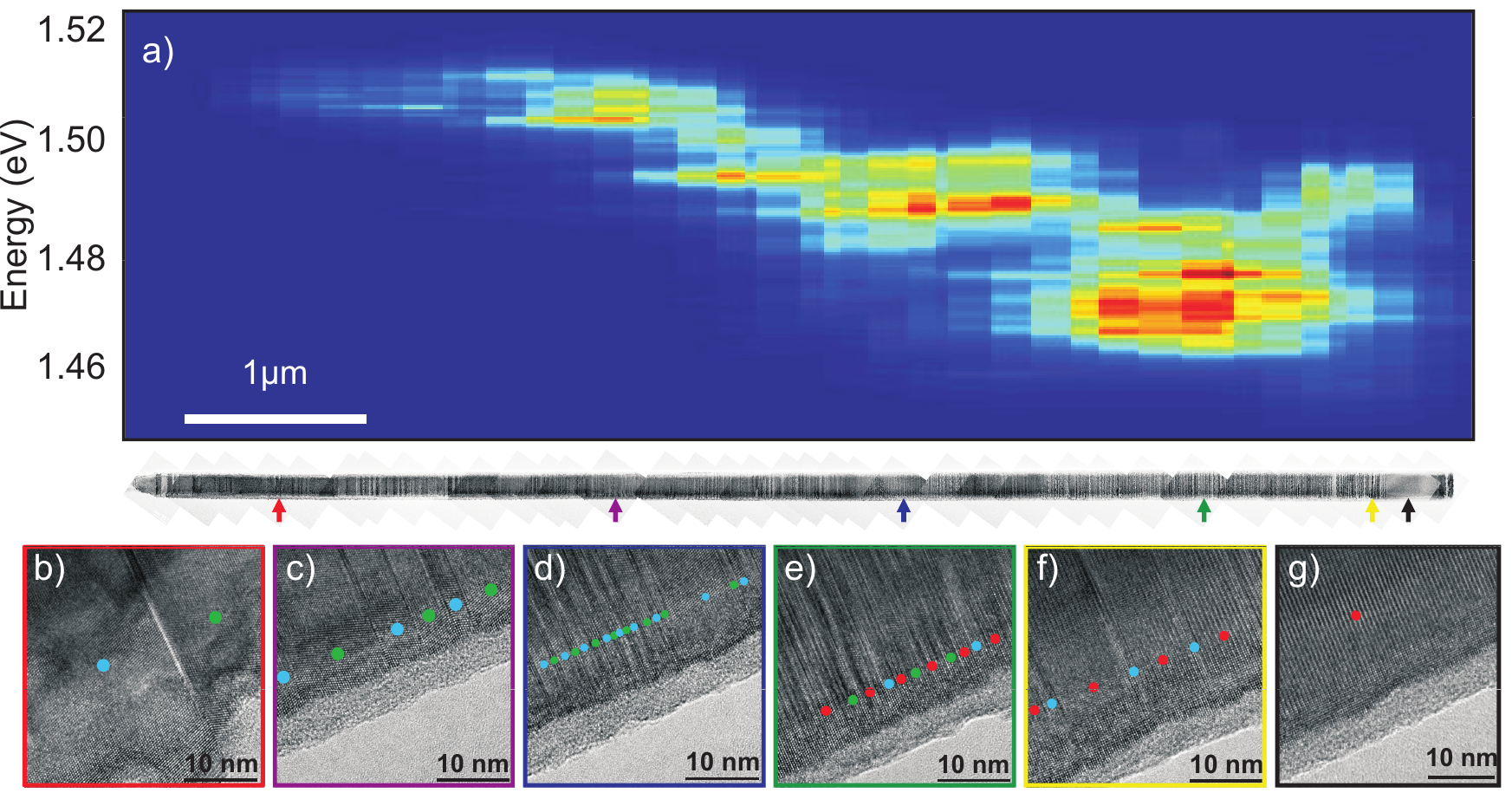}
\caption{\label{wzzbmappingcorr} a) Mapping of photoluminescence
spectra along the length of the GaAs wurtzite/zinc-blende
heterostructure nanowire shown below. The scale bar applies to both
the photoluminescence mapping and the TEM. b-g) HRTEM micrographs
showing the crystalline phases at the positions of the nanowire
indicated by the respectively colored arrows. Red dots mark the
wurtzite phase while blue and green dots show the zinc-blende phase. }
\end{figure*}

\begin{figure} \includegraphics[width=\columnwidth]{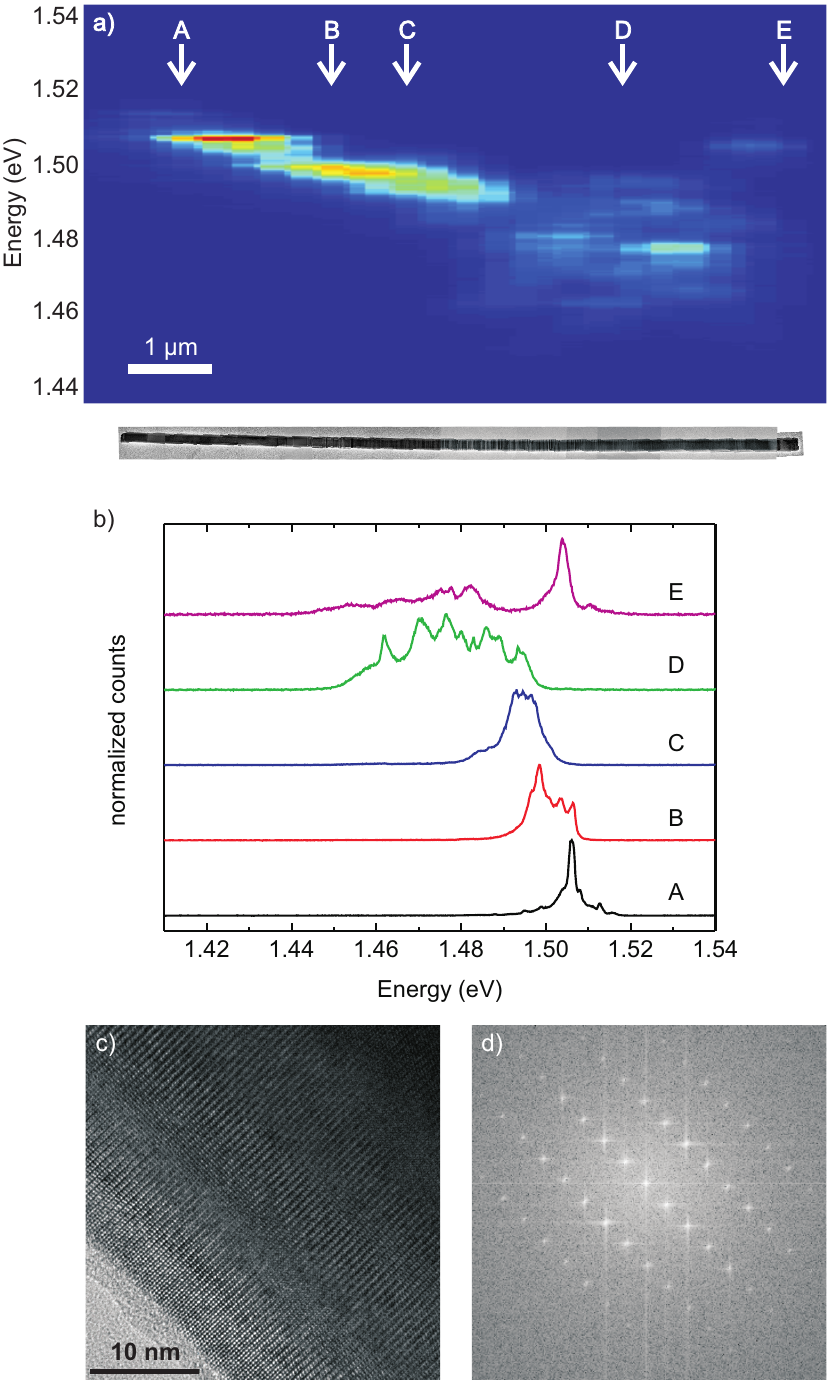}
\caption{\label{wzzbcorr2} a) Photoluminescence mapping and TEM
reconstruction of another wurtzite/zinc-blende heterostructure GaAs
nanowire from the sample. The scale bar applies to both the
photoluminescence mapping and the TEM. b) Individual photoluminescence
spectra of the five positions marked in a). All spectra are normalized
in intensity. c) HRTEM micrograph of region E of the nanowire. d)
Powerspectrum of micrograph c)} \end{figure}

\begin{figure} \includegraphics[width=\columnwidth]{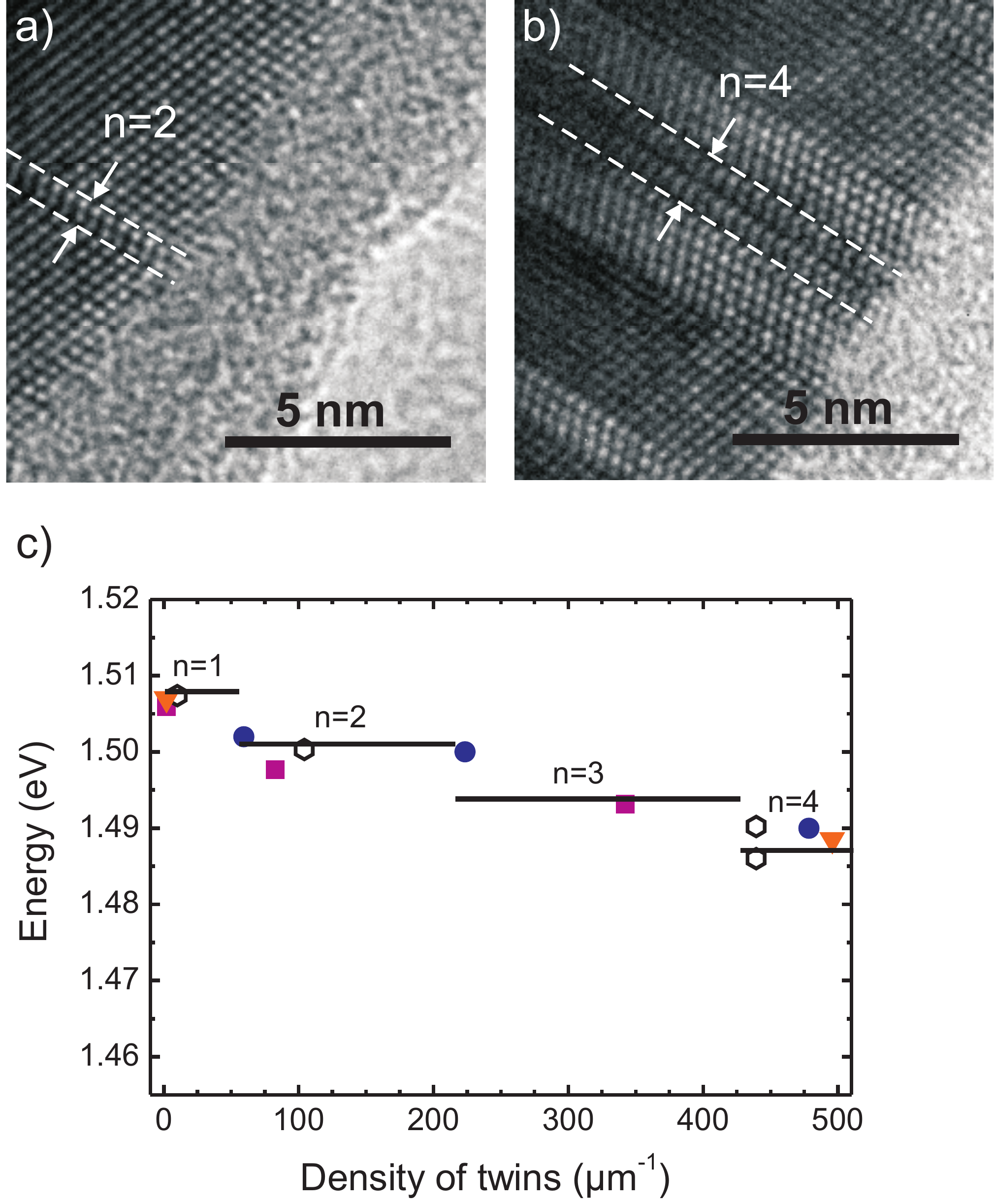}
\caption{\label{twindensity} HRTEM micrographs showing the occurrence
of: a) $n=2$ consecutive twins in a region with $<\,85$~twins/$\mu$m;
b) $n=4$ in a region with $>\,400$~twins/$\mu$m. c) Correlation of the
lowest PL transition energy to the density of rotational twins in the
corresponding part of the nanowire. The symbols show data obtained
from four different nanowires. The black line shows the transition
energies for wurtzite segments formed by $n = 1 - 4$ consecutive
twins. } \end{figure}

\begin{figure} \includegraphics[width=\columnwidth]{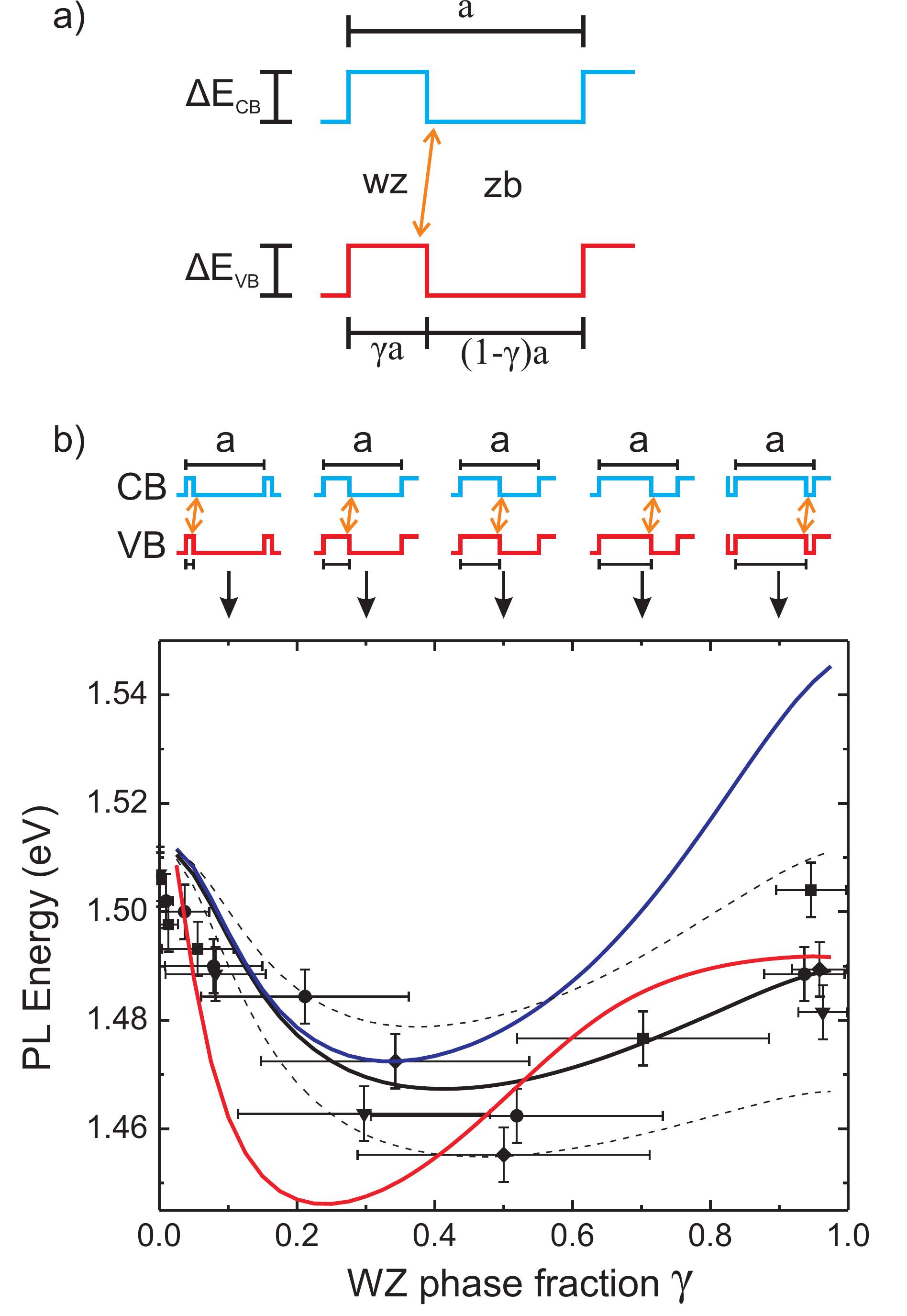}
\begin{tabular}{lccccl} & $\Delta E_{\mathrm{VB}}$ & $\Delta
E_{\mathrm{CB}}$ & $m^*_{h,{\mathrm{WZ}}}$ & $m^*_{e,{\mathrm{WZ}}}$ &
\\ & (meV) & (meV) & $(m_e)$ & $(m_e)$ & \\ \hline
{\color{blue}\rule[1ex]{20pt}{1.1pt}} & 84 & 117 & 0.51 & 0.067 &
Theory~\cite{Murayama1994} \\ {\color{black}\rule[1ex]{20pt}{1.1pt}} &
$76\pm12$ & $53\pm20$ & 0.51 & 0.067 & Optimized fit\\
{\color{red}\rule[1ex]{20pt}{1.1pt}} & 122 & 101 & 0.766 & 1.092 &
Theory (present work) \\ \end{tabular}

\caption{\label{superlatt} a) Schematic of the type II homogeneous
superlattice model with wurtzite phase fraction $\gamma$ and
periodicity $a=13$\,nm. b) Lowest PL transition energy as a function
of the fraction of wurtzite phase $\gamma$ in the corresponding part
of the nanowire (symbols). The colored lines show results from model
calculations. The dashed black lines give the prediction bounds of the
model fit to the experimental data.} \end{figure}

\begin{figure} \includegraphics[width=0.8\columnwidth]{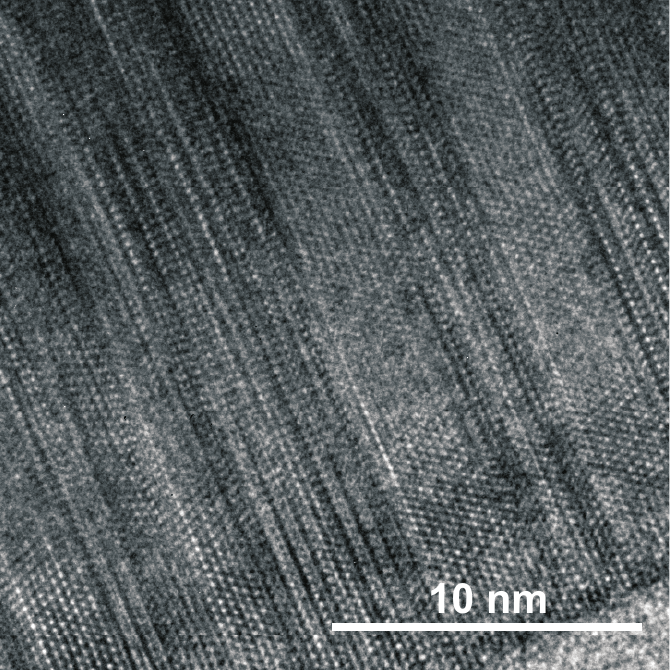}
\caption{\label{temlargesegment} HRTEM micrograph of the largest
segment of the nanowire with regions composed of equal fractions of
the zinc-blende and wurtzite phases, each approx. 6.5\,nm in length
along the nanowire axis.} \end{figure}

We now turn to the central aspect of this work, which is the
correlation between the structural and optical properties in
wurtzite/zinc-blende heterostructures. Representative SEM micrography
of the as-grown Ga catalyzed sample is shown in
Fig.\,\ref{semgacatalyzed}. The gallium droplet at the top of the
nanowire disappeared during the growth of the shell.

Fig.\,\ref{wzzbmappingcorr}a shows the $\mathrm{\mu PL}$ spatial
dependence in a 7\,$\mu$m long nanowire with a diameter of 170\,nm.
Below the PL scan we have included TEM micrographs representative of
various locations on the nanowire. From the structural point of view the nanowire is composed of two regions: (i) one with zinc-blende and a variation in twin density. (ii) one with a mixture of zinc-blende and wurtzite phases similar to a slightly disordered superlattice. The left side of the figure
corresponds to the tip of the nanowire, which is in the zinc-blende
phase with rotational twins separated by $\sim100$\,nm, shown in
Fig.\,\ref{wzzbmappingcorr}b. From left to right (towards the base of
the nanowire), the density of deviations from an ideal zinc-blende crystal structure increases in a regular way. 

Within 6\,$\mu$m from the left end of the nanowire these deviations are predominantly rotational twins and can be described considering a gradually increasing statistical twin density. For
example, at positions 1\,$\mu$m, 3\,$\mu$m and 5\,$\mu$m, we observe a
twin density of 60, 223 and 479~twins/$\mu$m, respectively, shown in
Fig.\,\ref{wzzbmappingcorr}c-d. This increase in the twin density is
accompanied by a regular red-shift of the PL peak from 1.50 to
1.48\,eV. The PL signal is spatially discontinuous.

In the final third of the nanowire the concept of twin density is not sufficient to describe the structure of the nanowire, as segments of wurtzite phase appear regularly. We can describe considering the fraction of wurtzite phase present in the nanowire.
At a distance of
6\,$\mu$m from the tip, the highly dense twin lattice turns into an alternation of
2 to 5\,nm thick zinc-blende and wurtzite sections. At that point, the
percentage of wurtzite is $\approx50\%$ and the average spacing of
each of the phases $\approx2$\,nm (Fig.\,\ref{wzzbmappingcorr}e). Moving forward towards the right,
the percentage of wurtzite quickly increases up to 100\%,
approximately 400\,nm before the bottom of the nanowire. Some stacking
faults and inclusions of zinc-blende smaller than 1\,nm are observed
right at the base, as shown in Fig.\,\ref{wzzbmappingcorr}f. A
continuous wurtzite region of 290\,nm is also found, shown in
Fig.\,\ref{wzzbmappingcorr}g. This region is followed by a single
2.4\,nm insertion of zinc-blende followed by 84\,nm of wurtzite.
Exactly at the base, we find a 46\,nm thick section with a mixture of
wurtzite and zinc-blende. The gradual decrease of wurtzite fraction
and twins is related to the increase in As$_4$ pressure during the
growth \cite{spirkoska-2009}.

Interestingly, for the nanowire section where the twin density
increases, the PL shifts from 1.51 down to 1.48\,eV. After this,
consistent with the regions where wurtzite sections appear clearly,
the PL peak shifts further down to 1.46\,eV. At the final section of
the nanowire a blue-shift of the PL energy is observed. We never
observe PL signal above 1.515\,eV. A total of four nanowires were
investigated with the direct correlation method. They all showed a
similar behavior. Another example is shown in Fig.\,\ref{wzzbcorr2}a.

The measurements presented above raise two main questions:\\ 1) In the
top two thirds of the nanowire, why is the twinning density shifting
the PL emission towards lower energies from 1.51 to 1.48\,eV?\\ 2) Can
one derive the values of the conduction and valence band discontinuity
between wurtzite and zinc-blende from the correspondence between the
TEM and PL measurements?

In order to address the first question, we start by analyzing what
seems to be a twin density dependence of the PL emission. Indeed, we
have observed a significant red-shift (40\,meV) for a twin density
varying from 50 to 500~twins/$\mu$m. Various causes may be responsible
for this. For example, it is known that the binding energy of excitons
confined in quantum well--like structures increases \cite{miller1}.
However in GaAs, even in the ideal 2D confinement limit, this could
not account for a shift of more than $\sim$12\,meV at most
\cite{miller1}. It is known that a change of effective mass within the quantum well plane could further increase the exciton binding energies~\cite{Ropers2007}. However the in-plane effective masses are not expected to change so significantly in wurtzite GaAs~\cite{De2010}.  Most importantly, the presumably type II nature of the
transition actually leads to a separation of electrons and holes thus
diminishing the Coulomb interaction between the carriers. Moreover, a
type II superlattice of single wurtzite layers in zinc-blende with
varying superlattice periodicity cannot be an explanation because it
would lead to a blue shift instead of a red shift. Finally, we must
consider the possible correlation between the twin density and the
appearance of single structures with several consecutive twins,
corresponding to thin sections of the wurtzite phase. The occurrence
of one twin is often illustrated as one monolayer of wurtzite.
Multiple consecutive twinning leads to the formation of thicker
wurtzite segments, which could account for the PL peaks below the GaAs
band gap.

For a relatively homogeneous twin density over a length of several
hundreds of nm, the twin density in that region gives directly the
probability $p$ for a twinning event to occur during nucleation of a
GaAs atomic layer in the (111) direction: 
\begin{equation}
p=\frac{a\rho}{\sqrt{3}} 
\end{equation} 
where
$a$ is the cubic GaAs lattice constant and $\rho$ is the density of twins in the region of the nanowire. Assuming statistical
independence of the twinning process~\cite{ref21} and the law of large
numbers the probability of $n$ subsequent twinning events to occur is given by
$p_n=p^{n}(1-p)$. We can therefore derive the average distance
between such occasional segments consisting of $n$ consecutive twins.

\begin{equation} d_n=\frac{a}{\sqrt{3}p_n} \end{equation} 
In order
to reveal the existence of multiple twinning and formation of thin
segments of wurtzite, we examined each of the HRTEM micrographs along
the nanowire. We found sections formed by $2-5$ monolayers (ML) of
wurtzite in regions with increasing twin density. HRTEM micrographs
corresponding to $n=2$ and $n=4$ are shown in
Fig.\,\ref{twindensity}a-b. The position of these sections corresponds
to regions in the nanowire with high PL intensity. In
Fig.\,\ref{twindensity}c we plot the PL peak position as a function of
the twin density of the nanowire section. Low temperature PL is very
sensitive to the presence of such inhomogeneities with lower
recombination energies. The majority of photogenerated carriers
diffuse to the wurtzite quantum wells (lower energy) where they
recombine. The line widths of the peaks are extremely sharp with a
FWHM below 2.5\,meV, which further corroborates our explanation
(Fig.\,\ref{wzzbcorr2}b). We also plot the position of the peaks
observed in positions where twin stacks with n between 1 and 4 are
observed by HRTEM. Clearly, the redshift in the PL peaks is correlated
with the existence of consecutive twins and the existence of 1, 2 and
3 ML thick single wurtzite regions.

In the last third of the nanowire, segments of wurtzite appear
regularly, as shown in Fig.\,\ref{wzzbmappingcorr}e. In the PL
mapping, multiple peaks are observed from 1.49 down to 1.455\,eV. The
final part on the right end of the nanowire is composed of pure
wurtzite GaAs material segments with a length of up to 290\,nm, see
Fig.\,\ref{wzzbmappingcorr}f-g, with small insertions of the
zinc-blende structure. This region shows again PL emission above
1.49\,eV. In order to address the question of whether the band offsets
between wurtzite and zinc-blende GaAs phases can be determined, we
plot the PL peaks as a function of the wurtzite phase fraction
$\gamma$ in Fig.\,\ref{superlatt}. The horizontal error bars represent the microscopic variations of the actual nonperiodic heterostructure from an ideal homogeneous superlattice. Peaks at 1.455\,eV correspond to
regions of the nanowire exhibiting approximately 50:50 composition of
the wurtzite and zinc-blende phases. Peaks at energies higher than
1.455\,eV correspond to sections in the nanowire richer in one of the
two phases. We point out that the highest energy PL peaks in the
region with pure wurtzite correspond to a band gap of 1.504\,eV; we
never observed PL at energies above the free exciton of zinc-blende
GaAs.

Finally, we propose a model to explain the PL emission in the last
third of the nanowire. To this end, we calculate the miniband
formation of a homogeneous superlattice (SL) with type II band
alignment and band offsets $\Delta E_{\mathrm{CB}}$ and $\Delta
E_{\mathrm{VB}}$, shown in Fig.\,\ref{superlatt}a, using a
Kronig-Penney model. The overall lattice periodicity was set to
$a_{\mathrm{SL}}=13$\,nm which is the widest periodicity observed by
TEM in the region of 50:50 phase composition, see
Fig.\,\ref{temlargesegment}. The largest periodicities in the structure correspond to the lowest recombination energies. Low temperature microphotoluminescence is most sensitive to the transitions with the lowest recombination energies. The calculations were performed for
various valence and conduction band offsets and are plotted in
Fig.\,\ref{superlatt}b varying the relative thickness of the wurtzite
and zinc-blende segments in the superlattice and considering the experimental band gap of zinc-blende GaAs in all cases. 
We want to note that the calculations are neglecting modifications of the band structure from strain \cite{zardo-2009} and spontaneous polarization of the hexagonal phase \cite{PhysRevB.56.R10024}.
As a first attempt to
provide a quantitative explanation, we use the values $\Delta
E_{\mathrm{CB}}=117$\,meV and $\Delta E_{\mathrm{VB}}=84$\,meV
obtained in earlier theoretical work~\cite{Murayama1994}, and
effective masses in wurtzite GaAs identical to the experimental ones
for zinc-blende GaAs. The result is shown in Fig.\,\ref{superlatt}b.
Clearly, there is a disagreement between the theory and the
experimental data for sections of the nanowire with a large wurtzite
content. By least square fitting of the experimental data with the
periodic superlattice model we can estimate $\Delta
E_{\mathrm{CB}}=53\pm20$\,meV and $\Delta
E_{\mathrm{VB}}=76\pm12$\,meV.

\subsection{Calculation of the band offset and effective masses}

\begin{table*}
\caption{\label{table:bands} The calculated energy band gaps and band
offsets: LDA and HSE06 refer to the present work (see text for
details). For comparison we also include the results of earlier work
using ab initio calculations \cite{Murayama1994,Yeh1994,Zanolli2007}
or an Empirical Pseudopotential method \cite{De2010}. Experimental
results for band offsets are from the optimal fit shown in
Fig.\,\ref{superlatt}. $\varepsilon_{\mathrm{gap-ZB}}$ is the band gap
of zinc-blende phase while $\Delta
\varepsilon_{\mathrm{gap}}=\varepsilon_{\mathrm{gap-WZ}}-\varepsilon_{\mathrm{gap-ZB}}$
corresponds to the difference in band gap between the zinc-blende and
wurtzite phases.} \begin{ruledtabular} \begin{tabular}{lcccc} &
$\varepsilon_{\mathrm{gap-ZB}}$ (eV) & $\Delta
\varepsilon_{\mathrm{gap}}$ (eV) & $\Delta E_{\mathrm{VB}}$ (meV) &
$\Delta E_{\mathrm{CB}}$ (meV) \\ \hline DFT-LDA~\cite{Murayama1994} &
0.614 &$+0.033$& 84 & 117 \\ DFT-LDA~\cite{Yeh1994} & 0.44 & $+0.03$&
- & - \\ DFT-GW~\cite{Zanolli2007} & 1.133 & $+0.218$& - & - \\
EP~\cite{De2010} &1.519\footnote{Based on fit to experimental zinc-blende band gap} &$-0.016$ & 79.2 & 63.2\\ 
DFT-LDA & 0.351
&$-0.044$ & 114 & 70 \\ DFT-HSE06 & 1.405 &$-0.021$ & 122 & 101 \\
experiment & 1.51 &$-0.023$ & $76\pm12$ & $53\pm20$ \\ \end{tabular}
\end{ruledtabular} \end{table*}

\begin{figure} \includegraphics[width=\columnwidth]{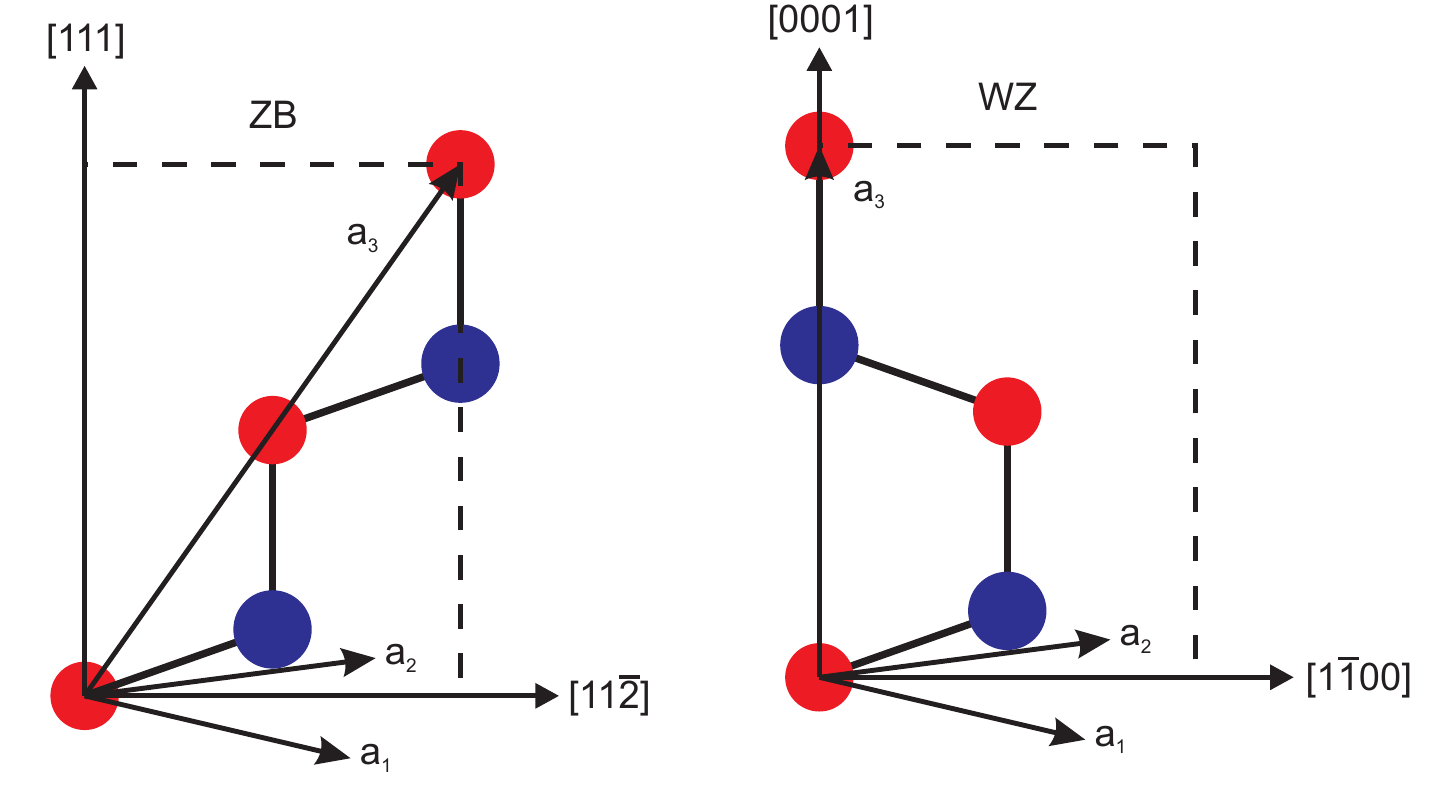}
\caption{\label{GaAs-structures} The structures used in the
calculations for the zinc-blende and wurtzite lattices, with lattice
vectors ${\bf a}_1, {\bf a}_2, {\bf a}_3$. The larger (blue) and
smaller (red) circles denote the two types of ions and the [111] and
[11$\overline{\text{2}}$] crystallographic directions in zinc-blende,
as well as the [0001] and [1$\overline{\text{1}}$00] crystallographic
direction in wurtzite are indicated.} \end{figure}

In order to obtain more accurate band offsets and effective mass
values, we performed electronic structure calculations in the context
of Density Functional Theory (DFT)~\cite{Kohn65} using {\sc
VASP}~\cite{Kresse96}. For the full band-structure throughout the
Brillouin Zone (BZ), we employed the standard Local Density
Approximation (LDA)~\cite{Ceperley1980,Perdew81}, with corresponding
pseudopotentials. It is well known that this type of approach does not
reproduce well the unoccupied part of the electronic spectrum and the
energy band gap. Recently, it was shown that both the band
gap~\cite{Marsman2008} and the excitation energies~\cite{Paier2008}
are vastly improved by applying a screened Hartree--Fock hybrid
exchange functional (HSE06)~\cite{Heyd2003,Krukau2006}. However, such
calculations due to their high computational cost can be performed
only for selected high-symmetry points in the BZ. Our approach then
consists of performing a full band-structure calculation with the LDA
method and using the HSE06 results to correct the excited states, by
applying a uniform shift equal to the band gap correction for the
zinc-blende structure; we are fully aware that such a ``scissor
operator'' approach may not be adequate~\cite{Remediakis1999}, but for
the bands of interest this approach gives very good results as
explained below. For the hybrid exchange calculations we employed the
PBE exchange-correlation functional in the DFT part~\cite{PBE} and PAW potentials to
represent the ions~\cite{Blochl2003}. We use a unit cell that has the
same number of atoms in the zinc-blende and wurtzite lattices, by
doubling one of the lattice vectors of the former, as shown in
Fig.\,\ref{GaAs-structures}. We used plane-wave energy cutoffs of
326\,eV for the LDA and 400\,eV for the HSE06 calculations, and a
reciprocal space Monkhorst-Pack grid of $8 \times 8 \times 4$
k-points~\cite{Monkhorst1976}. We used the experimental in-plane 
lattice constant of zinc-blende (3.99\,\AA) corresponding to bulk  
$a = 5.65$\,\AA. This is very close to the calculated ones, (3.95\,\AA)   for 
both the zinc-blende and wurtzite phases. For this we neglected the 
effect of strain and influence of band structure.

\begin{figure} \includegraphics[width=\columnwidth]{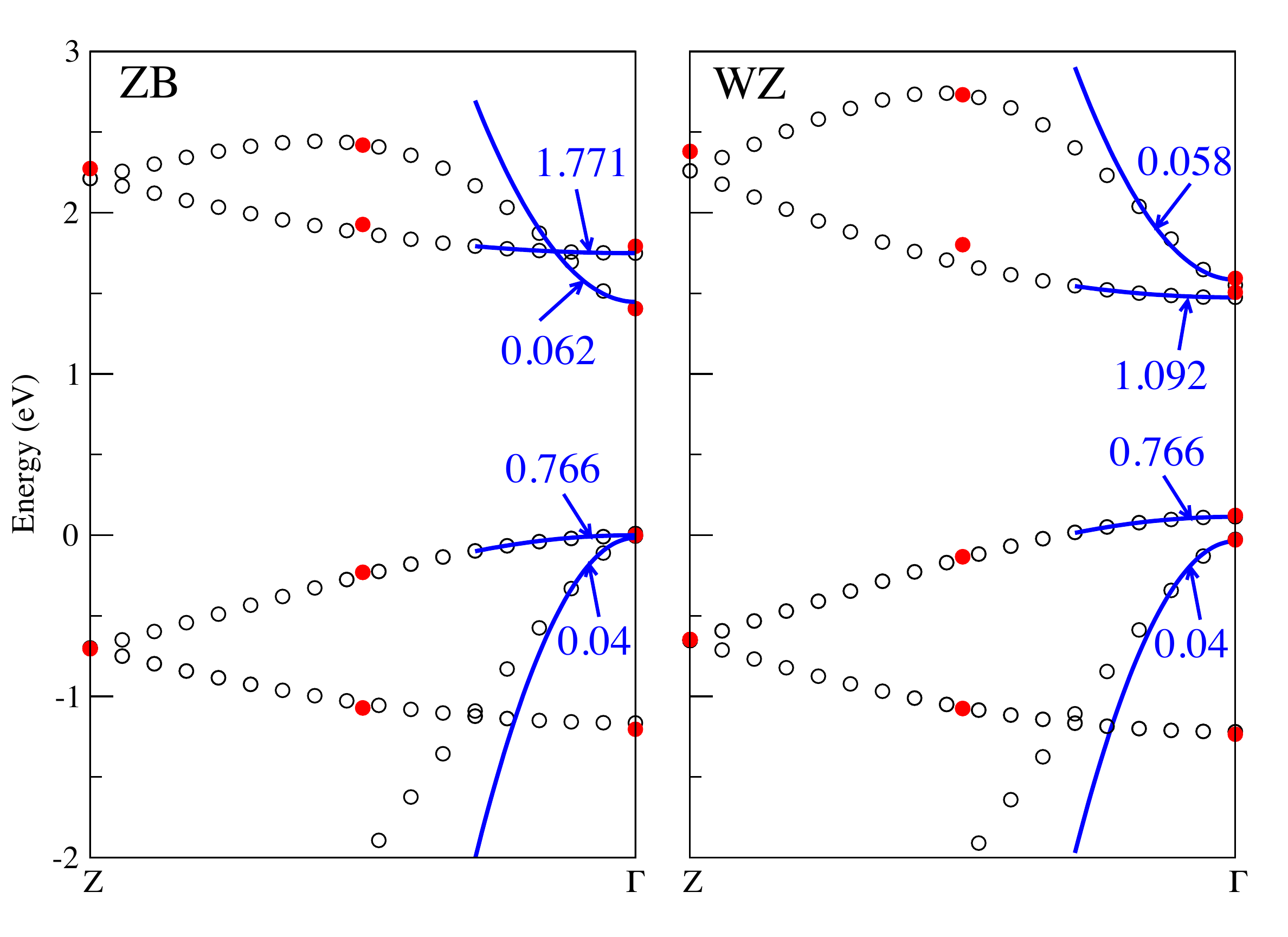}
\caption{\label{GaAs-bands} The band structures of the zinc-blende and
wurtzite lattices along the Z$\Gamma$ direction in the Brillouin Zone.
Open black circles are the results of the LDA calculations, red filled
circles are the results of HSE06 calculations and blue lines are
quadratic fits near $\Gamma$. The values in blue indicate the
effective masses in units of $m_e$ obtained from the quadratic fits.}
\end{figure}

The results of the calculations are collected in
Table\,\ref{table:bands} and the band structures for the zinc-blende
and wurtzite phases are shown in Fig.\,\ref{GaAs-bands}, along the
important directions, that is, [111] for zinc-blende and [0001] for
wurtzite. We can calculate the band structure of this interface and 
assign each band to relevant wurtzite or zinc-blende phase, based 
on density distribution and similarity to bands of the bulk zinc-blende 
and  wurtzite structure. In this way we can get information of band 
alignment/offsets of this interface. It is clear from the plot in 
Fig.\,\ref{GaAs-bands} that along these directions in
reciprocal space the uniform shift of excited LDA states (open black
circles) by an amount equal to the band gap correction for the
zinc-blende phase, brings them to excellent agreement with the
calculated HSE06 results (filled red circles). The calculated band
offsets are $\Delta E_{\mathrm{CB}}=101$\,meV and $\Delta
E_{\mathrm{VB}}=122$\,meV, and effective masses along the [0001]
wurtzite orientation are $m^*_{h,{\mathrm{WZ}}}=0.766 \; m_e$ and
$m^*_{e,{\mathrm{WZ}}}=1.092 \; m_e$; the corresponding values for the
[111] zinc-blende orientation are $m^*_{h,{\mathrm{ZB}}}=0.766 \; m_e$
and $m^*_{e,{\mathrm{ZB}}}=0.062 \; m_e$. The essential difference is
in the electron effective mass, which comes from the folding of the
bands along the [0001] wurtzite orientation, as seen in
Fig.\,\ref{GaAs-bands}. The corresponding superlattice model curve
using the calculated band offsets and wurtzite effective masses is
shown in Fig.\,\ref{superlatt}b. Our first principles calculations
indicate that the band gap of wurtzite GaAs is 21\,meV lower compared
to the band gap of the zinc-blende phase are in good agreement with
the experimental data. Our experimentally determined band offsets are
also in excellent agreement with band offsets recently calculated by
De and Pryor \cite{De2010} using an Empirical Pseudopotential method
(see Table\,\ref{table:bands}).

\section{Conclusions} In conclusion, we have presented results on a
new technique that enables the experimental direct correlation between
confocal micro-photoluminescence measurements and the structural
characterization with high resolution transmission electron microscopy
of nanowires consisting of nearly 100\% wurtzite and presenting
wurtzite/zinc-blende polytypism. The observations are in good
agreement with theoretical predictions for the band gap and band
offsets between the wurtzite and zinc-blende phases of GaAs. In
wurtzite GaAs nanowires, we observe photoluminescence always below
1.52\,eV. Taking into account the electronic confinement, the results
indicate that the electronic band gap of wurtzite GaAs phase is
slightly smaller than zinc-blende GaAs with an estimated
$\varepsilon_{\mathrm{gap-WZ}}=1.50\,\mathrm{eV}$.

The results on the polytypic nanowires are multiple. First, we
correlate the appearance of wurtzite sections with the probability of
twin formation. Then, we link the position of the luminescence with
the stacking sequence of the wurtzite and zinc-blende phases. Due to
the fact that the actual nanowire heterostructure is non periodic, an
exact calculation for the optical transitions would require a model
including the complete stacking sequence of each nanowire with atomic
precision; it is extremely challenging to determine this sequence
experimentally. Using a more simple periodic superlattice model, we
have demonstrated that the experimental findings can be well explained
despite the drastic simplifications of the model; we determine for the
first time an experimental estimation for the band offsets of $\Delta
E_{\mathrm{CB}}=53\pm20$\,meV and $\Delta
E_{\mathrm{VB}}=76\pm12$\,meV. These band offsets correspond to a band
gap for wurtzite GaAs of approx. 23\,meV lower compared to the band
gap of the zinc-blende phase. This is in excellent agreement with our
results on pure wurtzite GaAs nanowires, other
experiments~\cite{Moewe2008}, our first principles calculations and
other recent band structure calculations \cite{De2010} . Furthermore,
a slightly smaller band gap of the wurtzite phase of GaAs is supported
by our general observation that even though both nanowire samples
contain large amounts of wurtzite GaAs, we have never observed PL
emission above the band gap of zinc-blende GaAs.

\begin{acknowledgments} The authors thank G. Abstreiter, T. Garma, D.
Spirkoska and M. Bichler for their experimental support and
discussions. We thank A. Petroutchik and L. T. Baczewski for the
preparation of the Au-covered substrates for NW growth. This research
was supported by Marie Curie Excellence Grant 'SENFED', the DFG
excellence cluster Nanosystems Initiative Munich, as well as SFB 631
and ERC Starting Grant 'Upcon'. This work was partially supported by
the Spanish Government projects Consolider Ingenio 2010 CSD2009 00013
IMAGINE and CSD2009 00050 MULTICAT. J.A. acknowledge the funding from
the Spanish CSIC project NEAMAN and MICINN project MAT2010-15138
(COPEON). The authors would like to thank the TEM facilities in the
Interdisciplinary Center for Electron Microscopy in Lausanne and in
Barcelona. E.R. acknowledge financial support from the ERA Nanoscience
Project 'QOptInt'. E.R. and D.S. thank for support via SFB 689.
\end{acknowledgments}

\end{document}